\documentclass[preprint]{aastex}

\def\aa_{\aap}

\newlength{\bigfigwidth}
\begin{document}

\title{{\Large A contribution of $^{26}$Al to the O-Al anticorrelation\\
          in globular cluster red giants}}
\shorttitle{A contribution of $^{26}$Al to the O-Al anticorrelation}

\author{P.A.~Denissenkov\altaffilmark{1,2}, A.~Weiss\altaffilmark{2}}

\altaffiltext{1}{Sobolev Astronomical Institute, St.\,Petersburg University,
             Bibliotechnaya Pl. 2, 198504 St.\,Petersburg, Russia}
\altaffiltext{2}{Max-Planck-Institut f\"ur Astrophysik,
             Karl-Schwarzschild-Stra{\ss}e 1, D-85741
             Garching, Germany}

\vfill 

\begin{abstract}
We modify our combined (``deep mixing'' plus ``primordial'')
scenario explaining the star-to-star abundance variations
in globular cluster red giants in such a way that
it confirms with new experimental data:
{\it (i)} the new and better constrained (NACRE) thermonuclear reaction rates and 
{\it (ii)} a discovery of the O-Na anticorrelation in stars below
the main-sequence turn-off in the cluster NGC6752. For the latter we
propose that some main-sequence dwarfs in globular clusters 
accrete material lost by red giant primary components of
close binaries during a common envelope event. As a consequence of the new
reaction rates, we are drawn to the conclusion that  
the anomalies in [Al/Fe] in globular cluster red giants
are in fact manifestations of $^{26}$Al$^{\rm g}$ (instead
of the stable isotope $^{27}$Al) 
abundance variations produced by deep mixing.

\end{abstract}


\section{Introduction}

Several years ago we proposed
\citep{dea98}
a combined scenario
which explained self-consistently the star-to-star
abundance variations of C, N, O, Na, Al and Mg
in globular cluster red giants (GCRGs).
This scenario has two components: a ``deep mixing''
(or ``evolutionary'') and a ``primordial'' one. The deep mixing
component demonstrates that the overabundances of N, Na and
Al correlating with the deficiencies of C and O can be produced
inside the GCRGs themselves via the hydrogen-shell burning
in the concurrent CNO-, NeNa-, and MgAl-cycles.
It requires that some extra-mixing
bridges the outer wing of the H-burning shell (HBS)
with the base of the convective envelope (BCE). 
The decline of [C/Fe] with decreasing $M_{\rm V}$ in M92 
red giants reported for the first time by \citet{langer86} and
confirmed recently by \citet{bellman01} is convincing evidence of
deep mixing in GCRGs. Moreover, \citet{gratton00} have shown that
extra-mixing resulting 
in considerable changes of the surface C, N and Li abundances in
low-metallicity field red giants is rather the rule than
the exception. 

However, in order to reproduce the large ($ >1$\,dex) variations
of [Al/Fe]\footnote{$[{\rm A}/{\rm B}] \equiv
\log_{10}(N({\rm A})/N({\rm B}))_{\rm star} -
\log_{10}(N({\rm A})/N({\rm B}))_\odot$, where
$N({\rm A})$ is the number density of species A} and
the O-Al anticorrelation in
$\omega$~Cen red giants found by 
\citet{ndc95} we assumed  {\it (i)} that
the initial abundance of $^{25}$Mg had been increased
up to [$^{25}$Mg/Fe]\,=\,1.1 primordially, and
{\it (ii)} that the rate of the reaction
$^{26}$Al$^{\rm g}$(p,$\gamma)^{27}$Si, which
leads to $^{27}$Al via $^{27}$Si($\beta^+\nu)^{27}$Al,
was $10^3$ times faster than that given by
\citet{cf88}. The first assumption  -- 
being the primordial component of our combined scenario --
was supported by our accepted model of globular clusters'
self-enrichment. In this model (based on
the earlier ones of \citet{cayrel} and
\citet{bbt95}) type II supernovae
are considered to enrich globular clusters with
Fe and $\alpha$-elements (e.g. $^{16}$O and $^{24}$Mg), and
intermediate-mass ($M$\,=\,3-8\,$M_\odot$)
asymptotic giant branch stars
(hereafter, IM~AGB stars) with $^{25,26}$Mg and,
to a lesser extent, with Na and Al. The second assumption concerning
the $^{26}$Al$^{\rm g}$(p,$\gamma)^{27}$Si reaction
was permissible because the uncertainty factor in this
rate was as large as $\sim 10^3$
\citep{amc95} at that time, at least for typical
H-shell burning temperatures.

Since the publication of Paper~I the following important results
related to the problem of the origin of
the O-Na and O-Al anticorrelations in GCRGs have been obtained:
{\it (i)} the globular cluster self-enrichment model
has received new observational support \citep{jehin99,smith00}
and has been developed further by \citet{parmentier99};
{\it (ii)} detailed calculations of nucleosynthesis
in IM~AGB stars of low metallicities have confirmed
our earlier conclusion about their ability to produce
considerable amounts of $^{25}$Mg and $^{26}$Mg
\citep{lfc00,ben00};
{\it (iii)} a compilation of
new thermonuclear reaction rates
(NACRE) has been published \citep{nacre}; the new rate for
the $^{26}$Al$^{\rm g}$(p,$\gamma)^{27}$Si reaction is constrained
much tighter, 
with an upper limit of only $\sim$\,50 times that of
\citet{cf88} at the temperatures
relevant to the outer wing of
the HBS in GCRGs ($T$\,=\,40-50$\cdot 10^6$~K);
{\it (iv)} in addition to $\omega$~Cen the observations by
\citet{ivans99} provided a second cluster -- M4 --
for which four abundance correlations (the O-Na, O-Al, C-O and
C-N ones) have been observed simultaneously for a number of
red giants, and also in this cluster Al is found to be strongly
overabundant; 
{\it (v)} quite recently the O-Na anticorrelation
(and, possibly, also the O-Al one, both
seen before only in GCRGs) has been detected in stars
below the main-sequence turn-off (MSTO) in the cluster
NGC\,6752 \citep{gratton01}. \citet{ventura01} have
proposed an interpretation for these latter observations:
they have taken advantage of the globular
clusters' self-enrichment model and their finding that
at very low metallicity the so-called hot bottom burning,
i.e.\ thermonuclear processing at the BCE,
in IM~AGB models occurs at $T \geq 10^8$~K; \citet{ventura01}
emphasize that at such high temperatures O should be
depleted and Na enhanced, therefore they infer that the O-Na
anticorrelation in GCRGs most likely is 
a consequence of accretion of
material lost by IM~AGB stars of a previous generation by low-mass MS
stars (i.e.\ a pure primordial effect). Work is in preparation
(Denissenkov \& Weiss 2001, in preparation) to
examine this hypothesis with detailed models and within the framework
of our own approach for solving the abundance anomalies.

While the first two new results support the ideas of Paper~I, fact
{\it (iii)} is in clear contradiction to our original combined
scenario which required that the critical reaction rate for
$^{26}$Al$^{\rm g}$(p,$\gamma)^{27}$Si was set to the highest possible
value. Therefore, we modify our scenario in the present {\em Letter}
in such a way that it agrees with the new experimental results listed
above, for example that it can explain the observations concerning M4
as well. To anticipate our main result, we find that the Al-anomalies
observed cannot be understood in terms of the stable isotope
$^{27}$Al, but can be explained self-consistently and naturally within
the deep mixing scenario if they are due to $^{26}$Al (in ground state).

\section{Aluminium nucleosynthesis in the hydrogen shell of red giants}

In this section we will present new models concerning the abundance
anomalies in $\omega$~Cen and M4. The method and details of the
calculations (both of the evolutionary models and of the
nucleosynthesis) have been described elsewhere (e.g.\ in Paper~I) and
will not be repeated 
here. We just recall the important fact that the deep mixing and
nucleosynthesis calculations are done on background models obtained
from interpolation of the envelope and H-shell structure of several
selected sequences of full stellar evolution models. 
The models have mass ($M/M_\odot=0.80$) and initial chemical
composition appropriate for the two clusters ($Z=0.0005$ for
$\omega$~Cen and $0.001$ for M4). The deep mixing
is assumed to be of diffusive nature and is characterized by two
parameters: mixing depth and speed (expressed as a diffusion
constant). The difference to our earlier work is the use of the new
NACRE \citep{nacre} rates. Concerning the Al-nucleosynthesis, we fully
take into account the two states of $^{26}$Al (ground and metastable
state indicated by additional superscripts ``g'' resp.\ ``m'') and treat
them as two independent species in our network. 

The new NACRE-rate of the reaction $^{26}$Al$^{\rm g}$(p,$\gamma)^{27}$Si
is not fast enough for producing $^{27}$Al
(the $\beta^+$-decay product of $^{27}$Si) in GCRGs even if its upper
limit (approximately a factor 50 above the given rate) was used.
Almost all initial $^{25}$Mg, whose abundance is assumed
to be increased primordially, is transformed into
$^{26}$Al$^{\rm g}$ which in turn decays to $^{26}$Mg
rather than captures protons to produce $^{27}$Al. This implies that
our combined scenario presented in Paper~I is no longer feasible,
since it worked only due to the fact that it was possible at that time
to increase the proton capture rate by a factor of 1000.

\begin{figure*}[t]
\epsscale{0.7}
\plotone{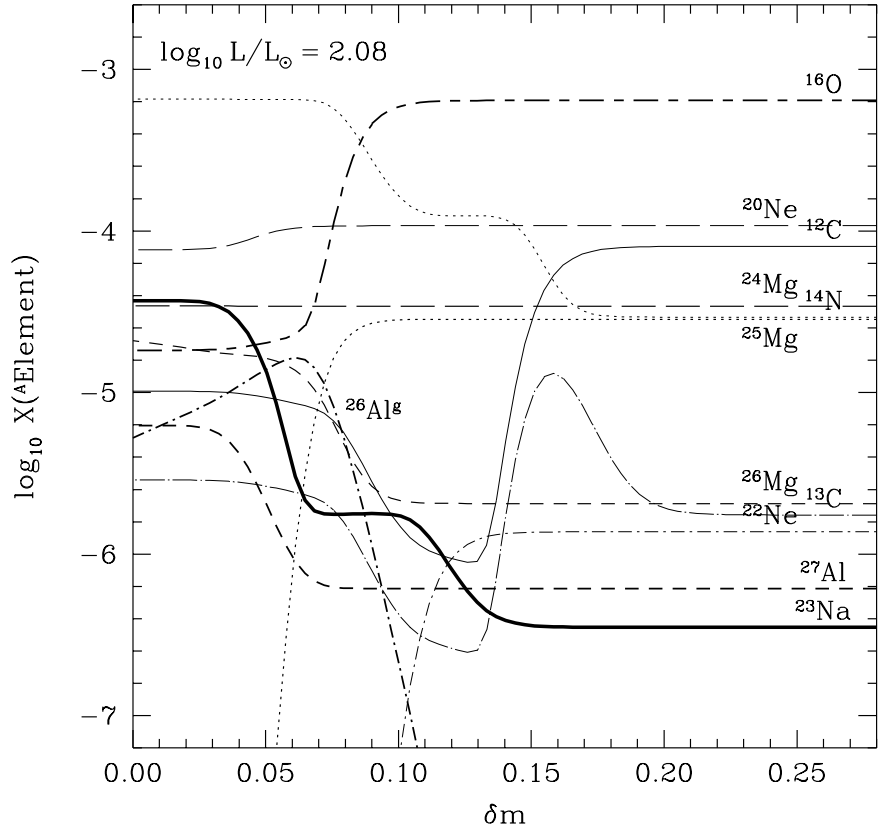}
\caption{Abundance profiles (logarithms of mass fractions) within the hydrogen shell of an
$0.8\, M_\odot$ model of luminosity $\log L/L_\odot = 2.08$ and
metallicity $Z=0.0005$ (or $\mathrm{[Fe/H]}=-1.58$), which corresponds to
the mean metallicity of red giants in $\omega$~Cen. For this model,
[Mg/Fe]=1.2 was assumed (primordial enrichment) and the standard
NACRE-rates were used. The abscissa is in our standard relative mass
coordinate $\delta m$, which is 0 at the bottom of the H-shell and 1
at the bottom of the convective envelope. 
\label{hshell}}
\end{figure*}

Figure~\ref{hshell} illustrates the abundance profiles due to
proton-capture nucleosynthesis within a red giant model typical for
$\omega$~Cen. We point out to the reader that 
the abundance of $^{26}$Al$^{\rm g}$ rises
steeply in layers farther outside the shell than does $^{27}$Al, with
a peak abundance almost half a magnitude larger. It is important to
emphasize that we used all NACRE reaction rates as published without
changing them within the possible error limits.

\clearpage

\begin{figure*}[t]
\epsscale{1.0}
\plotone{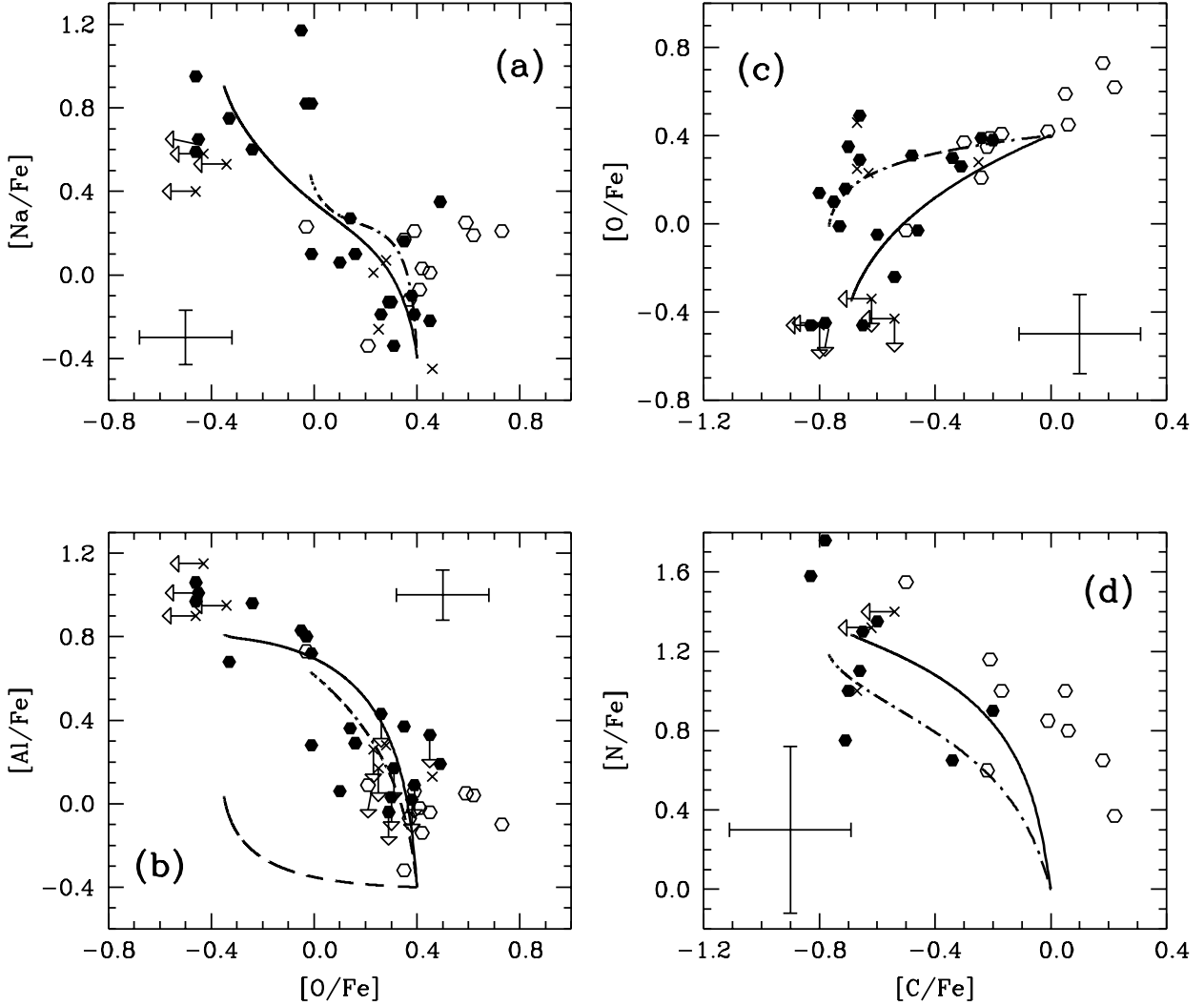}
\caption{The abundance trends seen in $\omega$~Cen red giants
                  (symbols; from Norris \& Da~Costa~1995)
                  compared with the results of our deep mixing
                  calculations. Two sets of mixing depth $\delta m$
                  (see text) and
                  rate (diffusion constant expressed in
                  cm$^2\cdot$s$^{-1}$) are used: ($\delta m_{\rm
                  mix}$;$D_{\rm mix}$) =
                  (0.05;\,$5\,10^8$) -- solid and dashed
                  lines --  and (0.065;\,$5\,10^8$) --
                  dot-dashed lines. The initial $^{25}$Mg abundance
                  was assumed to be [$^{25}$Mg/Fe]\,=\,1.2.
                  In panel b the solid and dot-dashed lines
                  show the pure $^{26}$Al$^{\rm g}$,
                  whereas the dashed line the pure $^{27}$Al abundance.
                  Large crosses indicate observational error bars.
                  Open and filled symbols refer to CO-strong and
                  CO-weak stars, and crosses denote stars with
                  unidentified CO status, following Norris \&
                  Da~Costa (1995).
                  \label{omcennacre}}
\end{figure*}

\begin{figure*}[t]
\epsscale{1.0}
\plotone{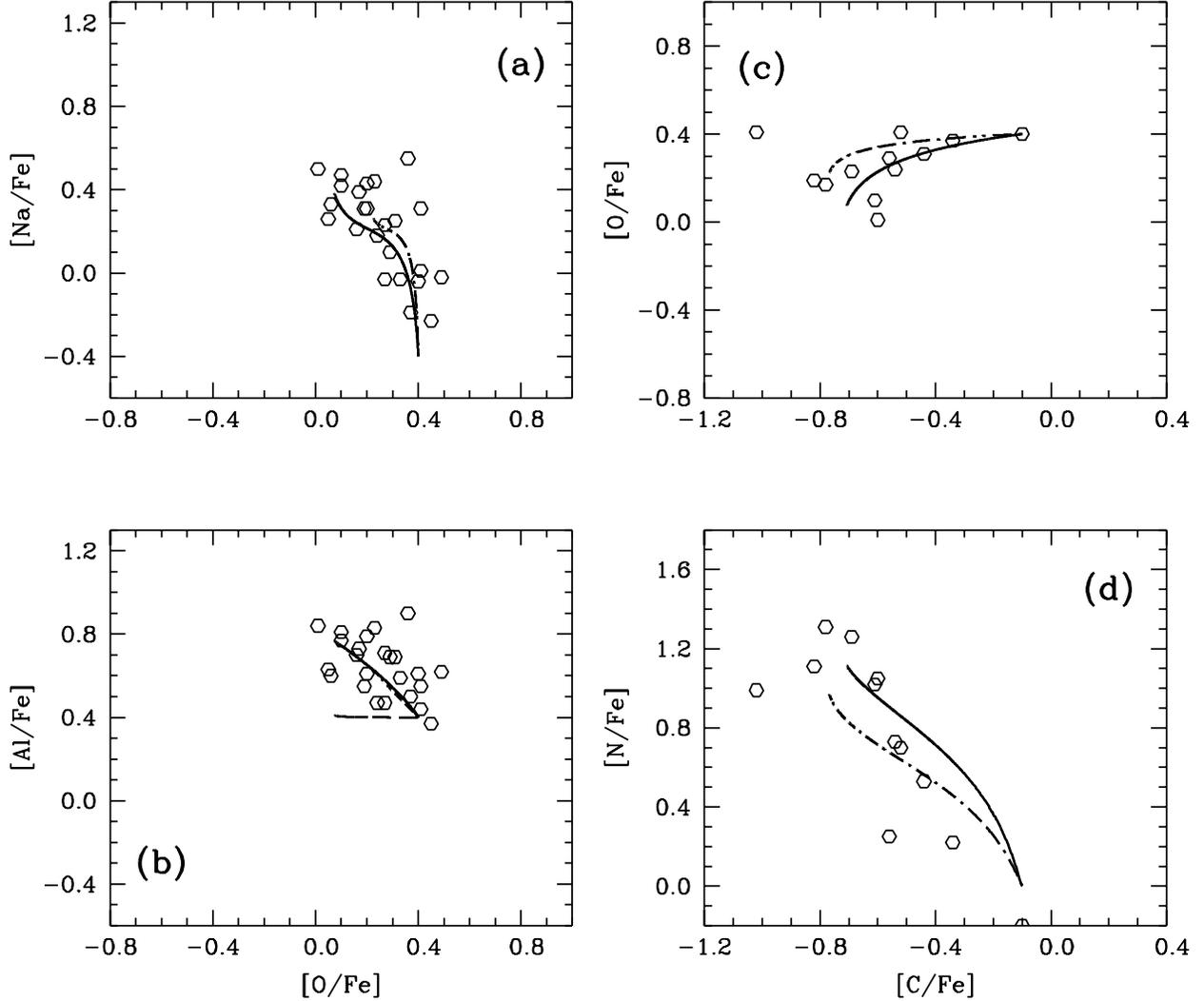}
\caption{Abundance trends seen in M4 red giants
                  (symbols; from Ivans et al.~1999)
                  compared with deep mixing results
                  for two sets of mixing depth and
                  rate ($\delta m_{\rm mix}$;$D_{\rm mix}$) =
                  (0.065;\,$4\,10^8$) -- solid and dashed
                  lines -- and  (0.075;\,$4\,10^8$) --
                  dot-dashed lines. The initial $^{12}$C, $^{25}$Mg
                  and $^{27}$Al abundances were assumed to be
                  [C/Fe]\,=\,$-0.1$, [$^{25}$Mg/Fe]\,=\,1.2 and
                  [Al/Fe]\,=\,$+0.4$, respectively.
                  Panel b,
                  solid line: pure $^{26}$Al$^{\rm g}$ yield;
                  dashed line: pure $^{27}$Al.
                  Calculations for this and Fig.~2
                  were done with the NACRE reaction rates
                  (Angulo et al. 1999). We emphasize the fact that
                  the mixing parameters used in Figs.~2 and 3
                  have very similar, but not identical
                  values.\label{m4nacre}} 
\end{figure*}

This offers the attractive  possibility to explain the
O-Al anticorrelation in GCRGs by assuming that the observed
variations of [Al/Fe] are in fact manifestations of the surface
$^{26}$Al$^{\rm g}$ abundance variations produced by deep mixing! 
In Figs.~\ref{omcennacre} and \ref{m4nacre} we illustrate that {\em
all known abundance variations and correlations} can be explained
simultaneously both for $\omega$~Cen and M4, with parameters for the
deep mixing which are similar but not identical for the two
clusters. It is also shown that the prediction for $^{27}$Al clearly
fails to reproduce the observations. Given this,
the question arises why the obvious possibility that the observed Al
is actually the isotope $^{26}$Al has not been
tried before? One reason might be that $^{26}$Al is unstable against
$\beta$-decay to $^{26}$Mg (emitting the famous 1.8~MeV $\gamma$-line)
with a life-time of $\approx 10^6$~yrs; therefore the survival of 
this isotope at the surface over the much longer lifetime of a red
giant might appear unlikely. However, within the deep mixing scenario
this argument does not hold necessarily.
Indeed, the following simple estimate supports our hypothesis:

The typical size of the radiative zone
between the HBS and
the BCE in GCRGs is $\Delta r$\,=1-2\,$R_\odot$. For a deep mixing
rate (a diffusion constant) of $D_{\rm mix}=4-5\cdot 10^8$
cm$^2\cdot$s$^{-1}$, with which the four abundance correlations
in the globular clusters $\omega$~Cen and M4 are
reproduced simultaneously (Figs.~2 and 3),
the extra-mixing turnover time
is $\tau_{\rm mix}\approx (\Delta r)^2/D_{\rm mix} = 0.3-1.5\cdot 10^6$
years. This is comparable with the life-time ($1.07\cdot 10^6$ years) of
$^{26}$Al$^{\rm g}$. Hence, part of the freshly synthesized
$^{26}$Al$^{\rm g}$ can survive the transport from the HBS to the BCE.
Our detailed deep mixing calculations confirm this simple estimate
(Figs.~2b and 3b). It should be noted that
in spite of the rather long life-time of stars on the RGB ($\sim 10^7$ years
for GCRGs) the surface $^{26}$Al$^{\rm g}$ abundance may be growing
steadily (its absolute value reaching
$X(^{26}{\rm Al}^{\rm g})$\,=\,9.4$\cdot$$10^{-6}$
in Fig.~2b, solid line)
because during all this time we have an active source of $^{26}$Al$^{\rm g}$
at the bottom of the deep mixing zone and assume continuing deep mixing.
Some increase of the $^{27}$Al abundance occurs only later on (Fig.~2b,
dashed line).

\clearpage
\clearpage

\clearpage

\section{The oxygen component of the O-Al anticorrelation}

The presence of s-process elements in
$\omega$~Cen and M4 red giants \citep{smith00,ivans99}
implies that AGB stars did contribute to the abundance anomalies 
seen in GCRGs. We have used this to argue for a primordial enhancement
of $^{25}$Mg in the models presented in the previous section.
Similarily, the correlating C, N, Na and O abundance
variations in stars below or around the MSTO in some globular
clusters \citep{briley96,gratton01} are apparent
signs of primordial pollution or accretion and not the result of deep
mixing, which we assume to happen only in evolved red giants. 

\citet{ventura01} have suggested that these anomalies are the result
of nucleosynthesis in and pollution by an earlier generation of IM~AGB
stars. We will investigate this idea elsewhere (Denissenkov \& Weiss
2001, in prepartion). Here we propose an alternative interpretation of
the O-Na (and O-Al) anticorrelation in stars below the MSTO in NGC6752
which does not preclude deep mixing in GCRGs but instead
takes advantage of it:
We assume that some MS dwarfs in globular
clusters might accrete material lost by red giant
primary components of close binary systems during the so-called
common envelope event \citep{il91}.
If these red giants had experienced deep mixing before
they filled their Roche-lobes then their ejected
envelopes could be enriched in Na and Al and be deficient in O.
Potentially any primary with 0.8$<$$M/M_\odot$$<$2.5 can play a r\^{o}le
in this scenario because in such stars the HBS erases
the molecular weight discontinuity left behind by the BCE
(which is thought to prevent
deep mixing from operating) before
these stars will reach the RGB tip.

\section{Conclusions}

Forced by new and better constrained proton-nucleosynthesis reaction
rates we modified our combined (primordial plus deep-mixing) scenario
\citep[Paper~I]{dea98} and reinvestigated the results of deep mixing
between the convective envelope and the hydrogen-burning shell in
GCRGs. We found that quite naturally, {\em all} (anti-)correlations in
$\omega$~Cen and M4 can be explained simultaneously if the observed
Al-isotope is in fact the unstable $^{26}$Al$^\mathrm{g}$ one. This is
possible only within the deep mixing scenario and in fact necessitates
ongoing exchange of matter between the $^{26}$Al$^\mathrm{g}$-source
in the shell and the convective envelope. As the primordial component
we have (as before) to assume an enrichment in $^{25}$Mg by pollution
with IM~AGB debris. Anomalies found around the MSTO clearly cannot be
explained within our scenario and require additional (primordial)
sources for which we tentatively suggest common-envelope effects in
close binaries, where the donor star has experienced deep mixing.

To verify our explanation for the Al-anomalies in GCRGs one could
either try to identify the isotope ratios of Al or make use of the
$\gamma$-line emitted during the decay of $^{26}$Al$^\mathrm{g}$
($^{26}$Al$^\mathrm{m}$ is present only in vanishing
amounts). Estimates have shown that the $\gamma$-flux from
$\omega$~Cen is expected to be of order
$10^{-7}$~photons/cm$^2\,\cdot$\,sec and therefore 2 orders of
magnitude below the COMPTEL detection limit, but might be within reach
of the planned ``Advanced Comptel Telescope''. However, we hope
that the ingenuity of observers might lead to the identification of
the Al-isotope to clarify this question.

\acknowledgements{
PAD wishes to express his gratitude for the warm hospitality
      to the staff of the Max-Plank-Institut f\"{u}r Astrophysik where
      this paper was prepared and acknowledges partial support of
      his work by the Russian Foundation for Basic Research (project
      code 00-15-96607). 
Valuable advice by R.~Diehl, A.~Iyudin and J.~Truran are
acknowledged. A.~Weiss thanks R.~Peterson and R.~Kraft for stimulating
discussions during a visit -- supported in
part by grants from the DAAD and the Fulbright Foundation --
to the Santa~Cruz astronomy department and for the kind hospitality he
received there.}

\newpage

\end{document}